# Dust Settling in Magnetorotationally-Driven Turbulent Discs II : The Pervasiveness of the Streaming Instability and its Consequences

By

David A. Tilley[1] (dtilley@nd.edu), Dinshaw S. Balsara[1] (dbalsara@nd.edu), Sean D. Brittain[2] (sbritt@clemson.edu), Terrence Rettig[1] (trettig@nd.edu)

**Running Head:** Dust Settling in Turbulent Discs II

**Mailing Address:**

1: Department of Physics
225 Nieuwland Science Hall
University of Notre Dame
Notre Dame, Indiana 46556
USA

2: Department of Physics
118 Kinard Laboratory
Clemson University
Clemson, South Carolina 29634
USA

**Phone:** (574) 631-2712

**Fax:** (574) 631-5952

**Abstract**

We present a series of simulations of turbulent stratified protostellar discs with the goal of characterizing the settling of dust throughout a minimum-mass solar nebula. We compare the evolution of both compact spherical grains, as well as highly fractal grains. Our simulations use a shearing-box formulation to study the evolution of dust grains locally within the disc, and collectively our simulations span the entire extent of a typical accretion disc. The dust is stirred by gas that undergoes MRI-driven turbulence. The turbulence tends to lift the dust above the disc's midplane, while gravity tends to draw it back to the midplane. This establishes a steady state scale height for the dust that is different for dust of different sizes. This sedimentation of dust is an important first step in planet formation and we predict that ALMA should be able to observationally verify its existence.

When significant sedimentation occurs, the dust will participate in a streaming instability that significantly enhances the dust density. We show that the streaming instability is pervasive at many of the outer radial stations in a disc. We characterize the scale heights of dust whose size ranges from a few microns all the way up to a few centimeters. We find that for spherical grains, a power-law relationship develops for the scale height with grain size, with a slope that is slightly steeper than -1/2. The sedimentation is strongest in the outer disc and increases for large grains. The results presented here show that direct measurements of grain settling can be made by ALMA and we present favorable conditions for observability. The streaming instability should also be directly observable and we provide conditions for directly observing it. We calculate collision rates and growth rates for the dust grains in our simulations of various sizes colliding with other grains, and find that these rates are significantly enhanced through the density enhancement arising from the streaming instability.

## 1) Introduction

Dust has many important roles in evolving protostellar discs. Dust regulates the heating and cooling of the gas in the disc through photoabsorption and shielding, and thus determines the temperature structure. As the disc evolves, its dust and gas distributions change, with concomitant evolution in the spectral energy distribution (SED) (Wood et al. 2002, Whitney et al. 2003). The evaluation of SEDs of discs becomes an important diagnostic of the distribution and nature of the dust in these discs (Wolf & Hillenbrand 2003), as well as the overall structure of young protostellar discs like HH 30 (Burrows et al. 1996). However the dust distributions for such SED modeling has always been derived from analytical models, such as Dubrulle, Morfill & Sterzik (1995) or Miyake & Nakagawa (1993). These models are based on a certain distribution of turbulence in the disc, and it has now become possible to test these distributions through the use of dynamical models of dust and gas (Balsara et al. 2009). Dust also plays an important role in building planets. The settling of dust to the mid-plane of an accretion disc and its resultant coagulation and growth plays an important role in planet formation (Safronov 1969, Goldreich & Ward 1973). A derivation of dust distributions that is obtained from a dynamical model is presented here and it is one of the goals of this paper to catalogue such information at several representative radii in a MMSN.

Dust is a key component in the process of planet formation. Observations of young stellar discs show that the removal of gas from the disc can occur in ~10 Myr, setting a limit on the time frame available for growing jovian planets (Sicilia-Aguilar et al. 2005). The core accretion model for the formation of gas giant planets postulates the initial formation of a rocky core on to which the gas can accrete. Mizuno (1980) and Pollack et al. (1996) show that radiative effects limit the rate of accretion in the core accretion model. Their calculations show that forming a gas giant planet requires more than a few million years. If coagulation and sedimentation occur, however, the opacity from dust could be significantly reduced and the time-scale for planetary core formation can be reduced significantly (Hubickyj et al. 2005). Planet migration within the disc could also accelerate the growth of planetary cores (Rice & Armitage 2003). Gravitational instability models for gas giant planet formation, which rely on collective processes in a massive disc acting on global scales, suggest that these planets can form on much smaller time-scales, as low as a few thousand years (Boss 1997, 2008). Furthermore, driving gravitational instabilities might require very massive discs, with masses that exceed several percent of the mass of the central star. Reconciling the core accretion and gravitational instability models with the observed disc masses and time-scales is thus a necessary step for progress in understanding planet formation. We explore here the evolution of dust within gaseous turbulent discs, with the goal of characterizing the time-scales for possible grain growth in such discs.

Coagulation can occur when dust grains collide and this process can grow dust into pebble and boulder-sized objects. These more massive particles are not as efficiently stirred by the turbulent motions in the gas, and can sediment into a thinner, more gravitationally unstable layer. Grain growth needs to be fast through this regime, as particles of these intermediate sizes are subject to a headwind-induced drag force that can lead them to spiral in to the protostar in a few hundred orbital time-scales (Weidenschilling 1977). Therefore, it is important to seek out rapid mechanisms for coagulation and grain growth. As coagulation is essentially proportional to the square of the density, the problem becomes one of raising the density sufficiently high to ensure that coagulation is fast. Whatever the mechanism for coagulation, the dust density in discs is low enough that it forms an essentially pressureless fluid. It will sediment to the midplane unless it can be stirred by turbulent motions transmitted to it by the gaseous component of the disc. Observations by Rettig et al. (2006), Bouwman et al. (2008), and Fedele et al. (2008) show evidence that dust does at least partially decouple from the gas and settle towards the midplane in accretion discs. Sedimentation will lead to greater dust densities in the midplane, and hence more rapid grain growth.

Dust densities can be further enhanced by significant amounts if it is subject to growing instabilities. Even in the absence of self-gravity, such instabilities do exist; in particular, the streaming instability of Youdin & Goodman (2005) can occur on fast time-scales if the density of the dust is comparable to the gas density in the midplane. The streaming instability develops when grains get entrained in the wake of nearby grains as they plough through the gas. This entrainment reduces the headwind that a grain experiences relative to the grains around it, leading it to move with a similar velocity. Thus, dust grains in close proximity to each other will start to move collectively. Initial studies of this phenomenon were performed by Youdin & Johansen (2007) and Johansen & Youdin (2007) in unstratified discs. Further studies by Johansen et al. (2007) extended these results to the case where dust self-gravity was incorporated, allowing the large dust grains to sediment into a thin layer within an unstratified gas disc, and then grow due to the combined effects of the streaming and gravitational instabilities. Balsara et al. (2009) examined the case with grains having a distribution of sizes in a stratified gaseous disc and found that the largest grain sizes can sediment to the midplane and undergo a streaming instability while smaller grains remain well mixed with the gas. They studied shearing sheet models that were stationed at two radii, 0.3 au and 10 au, in a MMSN. However, the streaming instability in this work only prevailed at one radius, 10 au. It is interesting to examine whether the instability persists at several radii in the disc, and if it becomes more vigorous at larger radii. In particular, the reduced gas densities in the outer regions of a protostellar disc should lead to stronger sedimentation in smaller grains. If the streaming instability can develop in these grains as well, it could serve as an even better precursor to planetesimal formation. Furthermore, it should be observable in the outer parts of these discs. Sedimentation of smaller grains would also be easier to observe and in this paper we show that ALMA could be well-positioned to observationally verify sedimentation in the outer parts of protostellar discs. A further goal of this paper is to show that the streaming instability is indeed quite pervasive in protostellar discs.

The shapes of dust grains in discs are not well-constrained by the data. Models and experiments of dust growth suggest that micron-sized dust grains might form a variety of fractal shapes (Ossenkopf 1993; Henning & Stognienko 1996, Blum et al. 2000) that reduce the mass and cross-section of the grains of a given fiducial size relative to a spherical grain. As the collision time-scale between gas and dust in the Epstein drag law regime is inversely proportional to the ratio of grain cross-section to grain mass, fractal dust grains are expected to couple with the gas differently than spherical grains. The BCCA (ballistic cluster-cluster aggregation) models of Ossenkopf (1993) are particularly interesting as they have particularly low volume filling fractions and a ratio of grain cross-section $\sigma$ to grain mass $m$ that varies with grain mass as $\sigma/m \propto m^{-0.05}$. As a result, large grains will couple almost as well to the gas as very small grains. These grains thus represent an extremely low degree of compactness (with spherical grains representing the other extreme, with $\sigma/m \propto m^{-1/3}$). One of the goals of this study is to understand the differences in the way spherical and fractal dust sediment in a protostellar disc. Absent observations, we simulate the settling of spherical grains and BCCA grains, which represent limits for the most compact and least compact grains we would expect to find in protostellar discs. By characterizing the behaviour of spherical and BCCA grains, we will be able to understand the role of fractal dust in sedimentation.

We present simulations of dust sedimentation in a minimum-mass solar nebula (MMSN) through shearing-box simulations performed at several radial stations around a 0.5 $M_\odot$ star. The disc is dynamically active and undergoes turbulence generated from the magnetorotational instability (MRI), with levels of turbulence that are consistent with observed accretion rates seen in T Tauri-type protostars. We incorporate an ensemble of dust grains with sizes ranging from 2.5 μm to to 2.5 cm, with mass ratios derived from the number distribution from Mathis, Rumpl & Nordsieck (1977, MRN henceforth) and a total dust-to-gas mass ratio of 0.01. We consider the role of both spherical and fractal dust grains. The structure of the turbulence has been described in Balsara et al. (2009) for standard dust-to-gas ratios at 10 au. In this paper, we extend our study to examine the structure of the dust throughout the disc from 0.3 au to 300 au, and to study the dynamics of MRI turbulence in gas-depleted discs.

In Section 2 we describe our simulation techniques, initial conditions and the parameter space we explore. Section 3 describes the results of our simulations and the consequences for developing the streaming instability throughout a disc. We summarize our conclusions in Section 4.

## 2) Description of Initial Conditions, Governing equations and Set-up of the Simulations.

We perform a suite of simulations in a shearing-box geometry to model the evolution of the gas and dust at several stations throughout a protostellar disc. We evolve the gas as a continuous fluid using the equations of ideal MHD. The particles are treated using a superparticle approximation, so that one computational superparticle represents an ensemble of dust grains. The forces arising from the interaction of these superparticles with the mesh-based fluid are interpolated to and from the underlying grid using triangular-shaped cell (TSC) interpolation. Absent their interactions with the gas, the particles are non-interacting and evolve ballistically.

The friction forces are added to the momentum equations of both the gas and dust, and take the form

$$\frac{\partial}{\partial t}\left(\rho_g \boldsymbol{v}_g\right) = \boldsymbol{f}_g - \sum_j \alpha_j \rho_j \rho_g \left(\boldsymbol{v}_g - \boldsymbol{v}_j\right) \tag{1}$$

$$\frac{d\boldsymbol{v}_j}{dt} = \boldsymbol{a}_j - \alpha_j \rho_g \left(\boldsymbol{v}_j - \boldsymbol{v}_g\right) \tag{2}$$

where $\rho_g$, $v_g$ and $f_g$ are the density, velocity and other forces acting on the gas, and $\rho_j$, $v_j$ and $a_j$ are the density, velocity, and sum of the other accelerations acting on dust particle of family ‘j’. Here, a family of dust is defined is defined as all of the dust of a particular grain radius. The summation in Equation (1) is over all of the families of dust. We use an implicit-explicit Runge-Kutta scheme to numerically treat the stiff source terms in Equations (1) and (2) that is described in detail in Balsara et al. (2009). The friction coefficient $\alpha_j$ described the strength of the momentum transfer between dust of family j and the gas. We characterize this friction coefficient through an Epstein drag law when the dust grain size $r_j < \frac{9}{4}\lambda_{\mathrm{MFP}}$ (Cuzzi, Dobrovolskis & Champney 1993), and a Stokes draw law when the grain size is larger than this. The form of the coefficient under these two drag laws, $\alpha_{\mathrm{Epstein}}$ and $\alpha_{\mathrm{Stokes}}$, are

$$\alpha_{\mathrm{Epstein}} = \frac{4\sigma_j}{3m_j}\sqrt{c_s^2 + \left|\boldsymbol{v}_j - \boldsymbol{v}_g\right|} \tag{3}$$

$$\alpha_{\mathrm{Stokes}} = \frac{3D}{8 r_j \rho_{\mathrm{grain}}}\left|\boldsymbol{v}_j - \boldsymbol{v}_g\right| \tag{4}$$

where $\sigma_j$ is the cross-section of the grain, $m_j$ is the mass of the grain, $c_s$ is the isothermal sound speed, $\rho_{grain}$ is the mass of the solid material from which the grain is made, and D is a coefficient that depends on a Reynolds number $\mathrm{Re}_{\mathrm{grain}} = 2r_j|\boldsymbol{v}_j - \boldsymbol{v}_g|/\nu$ of the gas moving past the dust grain ($\nu$ is kinematic viscosity of the gas):

$$D = \begin{cases} 24/\mathrm{Re}_{grain} & \text{if } \mathrm{Re}_{grain} \le 1 \\ 24/\mathrm{Re}_{grain}^{0.6} & \text{if } 1 < \mathrm{Re}_{grain} \le 800 \\ 0.44 & \text{if } \mathrm{Re}_{grain} > 800 \end{cases} \tag{5}$$

The mass and cross-section of the dust grains depend on whether the dust is spherical or fractal. For spherical grains, the cross-section and mass are $\sigma_j = \pi r_j^2$ and $m_j = \frac{4}{3}\pi\rho_{\mathrm{grain}} r_j^3$, respectively. For fractal grains, we use the BCCA models of Ossenkopf (1993), a particularly loose and fluffy fractal model that likely represents an extreme of grain non-compactness. In this model, if $r_j$ represents the extremal radius of the grain (i.e. the smallest radius that encloses the entire grain), then in the fractal limit the filling fraction of grain material within the extremal radius is $f_j = 0.279\left(r_j/0.01\ \mu\mathrm{m}\right)^{-1.04}\left(1 + 4.01\left[r_j/0.01\ \mu\mathrm{m}\right]^{-1.34}\right)$ (Henning & Stognienko 1996), where we have followed Henning & Stognienko (1996) and Suttner & Yorke (2001) in assuming that the constituent particles that make up the fractal grain are 0.01 μm in size. The mass of the grain is thus $m_j = f_j m_{\mathrm{sph}}$ and the cross-section is $\sigma_j = 0.692\sigma_{\mathrm{sph}}\left(r_j/0.01\ \mu\mathrm{m}\right)^{0.85} f_j^{0.95}\left(1 + 0.301/\left[\ln f_j + 3\ln\left(r_j/0.01\ \mu\mathrm{m}\right)\right]\right)$, where $m_{\mathrm{sph}}$ and $\sigma_{\mathrm{sph}}$ are the mass and cross-section of a spherical grain with the same extremal radius.

We perform three sets of eight simulations. Table 1 provides a list of simulation parameters for the minimum-mass solar nebula (MMSN) model. The first set of simulations evolves spherical dust grains in a MMSN, with a dust-to-gas mass ratio of 0.01. The second set considers BCCA fractal grains in a MMSN, also with a dust-to-gas mass ratio of 0.01. The final set considers spherical grains in a gas-depleted disc, such that the gas density has decreased by a factor of ten while the dust mass remains the same as in the first set. In other words, Σ is one-tenth of its values in Table I for this gas-depleted set of simulations. Thus, the dust-to-gas ratio in this suite of simulations is 0.1, and might represent a later transitional stage in the evolution of the disc.

As the shearing-box approximation requires that we simulate only a small portion of a disc, in each set of simulations we perform eight individual simulations at different radial locations spanning the entire likely range of a MMSN, from 0.3 au to 300 au. We initialize the simulations as described in Balsara et al. (2009). In the first two sets of simulations, we assume a disc surface mass density of $\Sigma(R) = \left(1700\ \mathrm{g\ cm^{-2}}\right)\left(R\ /\ 1\ \mathrm{au}\right)^{-1.5}$ and temperature distribution from D'Alessio et al. (1998). In the third set of simulations, we use $\Sigma(R) = \left(170\ \mathrm{g\ cm^{-2}}\right)\left(R\ /\ 1\ \mathrm{au}\right)^{-1.5}$, i.e. ten times lower. From the corresponding sound speed, assuming an isothermal equation of state, and a rotation rate $\Omega^2 = GM_* R^{-3}$, we obtain a scale height $\mathrm{H_{gas}} = \sqrt{2} c_s / \Omega$. We use this to derive a midplane gas density, $\rho_0 = \sqrt{\pi}\ \Sigma(R)\ /\ \mathrm{H_{gas}}$. Our initial conditions for the gas disc as a function of height above the midplane are Gaussian profiles $\rho_g = \rho_0 e^{-z^2/\mathrm{H_{gas}^2}}$. We initially distribute our dust particles within 2.5 gas scale heights of the midplane so that they have the same vertical profile as the gas density. We set the masses of the superparticles such that the total mass of the dust of all families is either 1% of the gas mass (sets 1 & 2) or 10% (set 3). The dust grains are divided into five dust families of different sizes from 2.5 μm to 2.5 cm (where "size" refers to either the radius of the spherical dust grain in Sets 1 & 3, or the extremal radius of the fractal dust grains in Set 2), and the relative fraction of the total dust mass in each family is given by an MRN distribution (Mathis, Rumpl & Nordsieck 1977).

In our simulations we do not include the effects of coagulation and growth, which take place over time-scales that are substantially longer than the ones simulated here. While Bouwman et al. (2003) have measured the grain distribution in HD 100546 and found a flatter distribution $N(r) \sim r^{-2}$, such a distribution represents the specific history of coagulation and growth in one system. As a result, the use of an MRN distribution in this paper represents but one possible conservative choice. Most of the qualitative results in this paper associated with the pervasiveness of the streaming instability would remain unchanged when different grain distributions are tried. Some of the quantitative results associated with collision times and mean free paths would depend sensitively on the dust distribution. However, the consequence of having a flatter distribution would be to collect a greater percentage of the dust mass in a few large particles. With more large dust grains, the collision times between large grains would be reduced. Consequently, flatter distributions of grains would only enhance our conclusions about vigorous grain growth in the presence of instabilities.

**3) Results**

The dust layers in our simulations experience sedimentation, as found in Balsara et al. (2009). As the density of these layers increase due to sedimentation, the dust layer begins to break up into clumps due to collective effects from the streaming instability. We show in Section 3.1 the role played by these collective instabilities in maintaining a scale height for the dust. In Section 3.2 we further examine the nature of this collective instability.

**3.1. Dust scale heights**

The thickness of the dust layer in a disc is related to the interplay between the turbulence in the dust and gravity. If the turbulence affects the dust as a diffusion process, and the dust is well-coupled to the gas, then balancing the diffusion rate with the sedimentation rate leads to an estimate for the ratio of dust and gas scale heights that varies as $H_d / H_{gas} \propto \tau_f^{-1/2}$ (Dubrulle, Morfill, & Sterzik 1995), where $\tau_f = (\alpha\rho_g)^{-1}$ is the friction time-scale between dust and gas. In the Epstein regime, the friction time-scale and grain size are related as $\tau_f \propto r_j$ for spherical grains and hence $H_d / H_{gas} \propto r_j^{-1/2}$. As a result, the thickness of the dust layers for grains of different sizes will vary with grain size. The estimate of Dubrulle, Morfill & Sterzik (1995) required several assumptions in order to make the problem analytically tractable. Firstly, they assumed that the gaseous component of the disc was in hydrostatic equilibrium, with no vertical motions, so that the role of gravity was to induce sedimentation at the terminal velocity of the grains. If the gaseous disc is turbulent as well, this assumption breaks down. Secondly, they assumed a particular $\alpha$-disc model for the turbulence in their calculation of the dust diffusion rate. Our goal here is to determine how the dust scale heights are set when we use 3D self-consistent simulations of turbulence driven via the MRI in the gaseous component of the disc, which in turn generates turbulence in the dust via frictional forces.

We can characterize the vertical distribution of dust by averaging the dust density in each x-y plane to create a vertical density profile. The Dubrulle, Morfill & Sterzik (1995) model predicts that the vertical distribution of dust remains Gaussian as sedimentation takes place. Previous numerical work has supported this hypothesis (Paper I). We can thus fit the Gaussian profile $\rho(z) = \rho_0 \mathrm{e}^{-z^2/H_d^2}$ to the vertical density profile in order to extract a scale height for the dust of each size. We note that both the form $e^{-z^2/H^2}$ and $e^{-z^2/2H^2}$ commonly appear in the literature; we have chosen the former. The scale heights calculated according to the former are larger by a factor of $\sqrt{2}$ than the latter. Generally, however, we will be making comparisons of the ratio of dust to gas scale heights, for which this factor of $\sqrt{2}$ cancels out.

We can use the scale height information to test whether our simulations have reached a steady-state balance between turbulent mixing and sedimentation. We plot the scale height $H_\mathrm{d}$ for the spherical grain simulations as a function of time in Figure 1. Sedimentation begins immediately in all of the simulations as the dust responds to the initial conditions, with the sedimentation occurring slowly for small (i.e. well-coupled) grains and quickly for larger grains. At the smaller radii, the smallest dust undergoes no sedimentation at all, i.e. $H_\mathrm{dust} \sim H_\mathrm{gas}$. At distances greater than 30 au, we also see that the largest dust grains are so poorly coupled to the gas that they undergo underdamped oscillations through the midplane. This behaviour is possible because our superparticle treatment of the dust allows streams of particles to interpenetrate; a continuum (mesh-based) representation of the dust would not show it. The oscillations of the 2.5 cm grains continue even at the end of our simulations at 100 au, 150 au and 300 au, as do the 2.5 mm grains at 300 au. The stopping times for these dust grains are several tens of orbits, and as a result we do not expect the to reach a steady state over the duration of these simulations. As our goal in this work is to characterize the dust once it has settled, we exclude those grain families that have not reached a steady state from our analysis. The dust scale heights of the other dust families do, however, become practically constant when averaged over one orbital time, and we consider these to have reached a steady state suitable for further analysis.

The scale heights for the BCCA fractal dust are quite different, as shown in Figure 2. The fractal grains of different sizes have nearly identical scale heights everywhere throughout the disc, as the ratio of cross-sectional area to mass is nearly constant with grain size. As a result, grains of different sizes experience nearly identical accelerations. This is related to the fact that $\sigma/m \propto m^{-0.05}$ for the fractal dust adopted here. Furthermore, the low volume filling fractions for the fractal grains result in grain masses that are ~4 x $10^{-8}$ times that for spherical grains at centimeter sizes. As a result, the gas-dust friction forces (which are reduced by a factor of only ~3 x $10^{-2}$ for fractal grains) are more than sufficient to keep the dust well-mixed with the gas. The one exception to the concordance between BCCA fractal grains of different sizes is for the 2.5 cm grains at 0.3 au, which have experienced some degree of sedimentation. Examining the drag law (Equations 3 & 4), we see that these grains have radii larger than the mean free path in the gas in the disc midplane, and as a result are in the Stokes rather than Epstein regime. This change in the character of the drag law allows these large grains to sediment somewhat. Since the Stokes law yields a smaller force per unit mass, the fractal grains at 0.3 au undergo sedimentation. It should be noted that the spherical 2.5 cm grains at 0.3 au are also in the Stokes regime, as our criterion for the transition from Epstein to Stokes is based on a comparison of grain radius to gas mean free path, and does not depend on the mass of the grain.

Our choices of spherical and BCCA grains are essentially the most extreme grain shapes we could have chosen, in terms of compactness versus non-compactness. Other fractal grain models tend to be somewhere within this range. BPCA fractal models (Ossenkopf 1993; Henning & Stognienko 1996), for instance, are significantly more centrally condensed than BCCA grains and have masses that scale with radius similarly to spherical grains, but with a filling fraction of ~4.6%. Such grains would thus be expected to behave intermediately to the spherical and BCCA grains we discuss here. Our simulations thus bound the dust scale heights that are likely achieved in protostellar discs.

In Figure 3a we plot the scale height of the dust in physical units as a function of radius in the disc after it has reached the steady state seen in Figures 1 & 2. The scale height of the gas is shown by the solid black line and provides an upper bound to the scale height of the dust. The spherical grain scale heights (Figure 3a) are well-mixed with the gas and closely follow the gas scale heights in the inner disc, a trend that can also be seen in Figures 1. At a distance of a few au, the largest (2.5 cm) grains begin to sediment out, developing into an almost flat profile with increasing radius. Successively smaller families of dust do show strong sedimentation at larger radii. The greater sedimentation in the outer disc arises from two factors; the increased dust-gas friction time-scale due to the reduced gas density in the outer disc, leading to poor coupling; and the weaker turbulence available to stir the dust layer away from the midplane, due to the decreased sound speed from the lower temperature in the disc. We note that as we use a fixed resolution of 192 zones in the z-direction and a physical extent of 3 $H_{\mathrm{gas}}$ in the z-direction, the resolution of our simulations imposes a minimum scale height that we can resolve, $2\,\Delta z_{\mathrm{midplane}}$, that increases with the radius in the disc. This resolution limit is shown with a dashed line in Figure 3, and we can see that the 2.5 cm grains at 30 au and the 2.5 mm grains at 150 au follow the resolution limit, instead of the overall trend. As a result, these points have been drawn with a dotted line.

We also show in Figure 3b the scale heights for a disc with the same initial conditions as the minimum-mass solar nebula, but in which the gas density has been decreased by a factor of 10. The scale heights have been substantially decreased for all grains as the momentum available to stir the dust has been reduced. The mass of dust in these simulations is the same as those for the minimum-mass solar nebula discussed previously, and so the smaller scale heights in turn lead to an increased dust density in the midplane. We will show later that this has important consequences for the possibility of grain growth in later stages of disc evolution.

The differences in the degree of sedimentation between the dust of different sizes will have important consequences for the prospects of observing dust in protostellar systems. We can use the scale heights in Figure 3 to interpolate the scale height of dust of a particular size at any point in the disc, using a linear interpolation method in log (r) and log ($H_d$). Similarly, we can calculate the density of dust in the midplane as a function of radius in each of our simulations, and interpolate those midplane densities to any point in the disc. With these interpolated scale heights and midplane densities for dust of different sizes as a function of radius, we can construct the global density structure of the disc in the log r – log z plane. This is shown in Figure 4, along with the minimum resolution of our data. Note that the axes are logarithmically scaled.

The density structure of the small, 2.5 μm dust closely follows that of the gas. Similarly, the 25 μm grains closely follow the gas density in the inner parts of the disc, but start to sediment out beyond 100 au. Sedimentation begins to appear in the 250 μm grains at ~30 au, and drops precipitously at 150 au. The 25 μm and 250 μm grains have become clearly separated from the 2.5 μm grains in the outer disc, and we predict that the distinction in their scales heights could be observable with high-resolution instruments viewing edge-on discs. The ALMA 350 μm and 1 mm bands, for instance, would be able to resolve structures larger than ~0.7 au and 2.1 au respectively at a distance of 140 pc, the distance of the Taurus molecular cloud. The 250 μm grains beyond 30 au and the 2.5 mm grains beyond 3 au have sedimented into a flat sub-disc, with virtually no dust of this size to be found beyond 5 au and 1 au respectively above the midplane. Similarly, in the regime in which we can resolve the 2.5 cm dust grains, we see that all of the 2.5 cm dust is contained within ~0.1 au of the midplane, and the flatness of this dusty sub-disc in the region of the disc that our simulations can adequately resolve suggest that little 2.5 cm dust would be found higher above the midplane than this further out in the disc. Since this sedimentation of dust is an important first step in planet formation, we believe that an observational demonstration of sedimentation would be a very important discovery. The simulations also show that one is most likely to make that discovery in the outer parts of discs. *By identifying the dust sizes that sediment at different radii, the simulations even help observers in identifying the wavebands in which the observations of dust settling need to be made.*

We are now in a position to test the dependence of the scale height with grain size. In Figure 5a we plot the ratio of the dust scale height of spherical grains to the gas scale height as a function of grain size for our simulations at each radial station. We see that the large grains sediment significantly, forming an approximate power law relationship between scale height and grain size. This is most pronounced in the outer disc, where the slope is the steepest. In the inner disc, the degree of sedimentation is quite small, and we see a much shallower slope. The dust scale height is limited by the gas scale height (Dubrulle, Morfill & Sterzik 1995), and the flattening we see in the slope of the dust scale heights when it becomes large is consistent with this expectation. These trends generalize consistently with the specific results at 10 au that we presented previously in Balsara et al. (2009). Furthermore, Fromang & Nelson (2009) have reported that $H_{\rm dust}/H_{\rm gas} \propto r^{-0.2}$ for dust grains in the size range of 1-100 μm at orbital radii of 1-8 au, consistent with the flattening of the slopes for smaller grain sizes that our results here produce.

The fractal grains, by contrast, are shown in Figure 5b. These scale heights are essentially independent of grain size. Sedimentation only occurs to any significant degree at 0.3 au for 2.5 cm grains, which are large enough to be in the Stokes drag law regime and hence feel a smaller frictional force than the other dust grains in the Epstein drag regime.

The spherical grains in Figure 5a at different radii are in very different environments, due to the overall structure of the disc. In Figure 5c we plot the spherical grain scale heights as a function of the dimensionless stopping time $\Omega\tau_s$. The grains at all radii now fall on the same curve, and we can clearly identify a power-law slope for large stopping times (and hence, large grain sizes). We fit the data to the semi-analytic curves of Dubrulle, Morfill & Sterzik (1995),

$$H_{dust}/H_{gas} = h\left(1+h^2\right)^{-1/2} \qquad \text{where} \qquad h^2 = (3/8)^{1/2}\,\alpha/\Omega\tau_s$$

where α is a parameter that measures the turbulent stresses. In Paper I we calculated this value explicitly from our data, and found that $\alpha \sim 2.3\text{x}10^{-3}$. Here, we determine α from a fit to our data, and find $\alpha=2.35\text{x}10^{-3}$. Our simulations simulations produce a slightly steeper power law relation between scale height and stopping time than the predicted slope of $(\Omega\tau_s)^{-1/2}$ from several semi-analytical models that consider the balance between sedimentation and turbulent diffusion (Dubrulle, Morfill & Sterzik 1995; Fromang & Papaloizou 2006), although the fit produces a value for α that is a good match to the turbulent stresses that we measure in our code.

### 3.2. Existence of the streaming instability

Youdin & Goodman (2005) demonstrated that when a dust layer couples marginally to the gas and is sufficiently dense, it can become unstable in the presence of a steady drift with respect to the gas due to a collective instability they dubbed a streaming instability. Specifically, the growth rates of this instability in the linear regime are strongly dependent on both the ratio of the density of dust to the density of gas, and the ratio of the dust-gas friction time-scale to the orbital time-scale. The growth rate becomes largest when the dust-to-gas density ratio becomes unity. Furthermore, it is known to increase with $\Omega\tau_s$. The Youdin & Goodman (2005) linear analysis relied on the sub-Keplerian rotational velocity of the gas to generate a constant initial relative velocity between the dust and gas in order to drive the instability. However, similar collective effects have been observed in simulations where the dust-gas relative velocities arise from turbulence generated from Kelvin-Helmholtz instabilities (Johansen, Henning & Klahr 2006) or through the MRI (Johansen et al. 2007, Paper I). Furthermore, in Paper I we presented evidence that while the dust in our simulation at 10 au become unstable due to this effect, no corresponding instability appeared at 0.3 au. We know from Figure 3 that sedimentation is strongest in the outer disc, and hence the dust-to-gas ratio in the midplane will be larger further away from the protostar. It is therefore of interest to ask where one would expect the streaming instability to operate in the disc.

To estimate the degree to which the streaming instability could operate in our disc models, we plot in Figure 6 the maximum dust-to-gas ratio in the midplane as well as the mean dust-to-gas ratio in the midplane for spherical dust grains. We see that while the mean dust-to-gas ratio for the largest dust can increase significantly over its starting value from the initial MRN distribution, it remains below 0.1 at all radii for all of the dust families with resolved scale heights. The dust layers that collapse to an unresolved thickness would likely have greater dust-to-gas ratios than are observed here, but this data are not shown in Figure 6. The maximum dust-to-gas ratios in the midplane at each radial station is ~1 for $R \geq 30$ au, so local enhancements in the dust density can bring the dust-to-gas ratio up high enough that collective effects could begin.

In Figure 7 we plot the maximum and mean dust-to-gas ratios in the midplane for the fractal dust. Here, very little sedimentation has occurred, and as a result the dust-to-gas ratio is always very tiny. As a result, we do not expect to see the streaming instability operate in these simulations, and indeed we do not.

The other important quantity Youdin & Goodman (2005) found that controls the growth rate of the streaming instability is the dimensionless stopping time $\Omega\tau_{\mathrm{f}}$. We plot this quantity for both spherical and fractal dust grains in Figure 8. In making Figure 8, in order to determine whether the grain is in the Epstein or Stokes regime, we assumed that the relative dust-gas velocity was 0.1 times the sound speed, which is the typical velocity dispersion in our simulations. We clearly set the transition from Epstein to Stokes drag laws in the dust at 0.3 au and 1 au.

As we would expect, the dimensionless stopping time is maximized for large dust and in the outer disc. The 2.5 cm grains begin to decouple ($\Omega\tau_f > 0.1$) at 10 au, and fully decouple ($\Omega\tau_f > 1$, indicating that the particle motions are becoming rather ballistic over the course of an orbit) at 100 au. Similarly, the 2.5 mm grains begin to decouple at 100 au, as do the 250 μm grains at 300 au. The streaming instability is strongest for $\Omega\tau_f \sim 1$, at least in the case when the instability is being driven by pressure gradients in the gas and hence the time-scales for coupling and driving are balanced, and our results indicate that a similar results hold for MRI-driven turbulence.

For the fractal dust in Figure 8b, the stopping times remain very small due to the fact that the grain cross-section to mass ratio is nearly constant with grain size. Here too, though, we see the transition from Epstein to Stokes drag at 0.3 au and 1 au. The fractal dust never develops $\Omega\tau_f \sim 1$, and so this avenue to promoting the streaming instability is not available, either.

In Figure 9 we show the total dust density of large grains (i.e. the sum of the 250 μm, 2.5 mm and 2.5 cm grain densities), in order to ascertain where streaming instabilities might be developing. We show this for the minimum-mass solar nebula in the upper set of eight plots, and for a nebula in which the gas has been depleted by a factor of 10 in the lower set of eight plots. For the minimum-mass solar nebula, we see very little density variations in the midplane in the inner disc, from 0.3-3 au, suggesting that no streaming instability is occurring within this region and hence the only density enhancement is from sedimentation. At 10 au and beyond, we do see increasingly large amounts of dust collecting in small volumes. In Paper I, we showed that at 10 au, this was the result of the streaming instability operating in the midplane for the 2.5 cm dust grains. Further out in the disc, the condensations we see arise from clumpings of progressively smaller dust grains. Since the ALMA instrument favors observation of smaller grains and can resolve features of ~5 milliarcseconds in its 350 μm band (or ~0.7 au at a distance of 140 pc), it suggests that the au-scale condensations of 250 μm dust that we see develop at 300 au are resolvable by ALMA at the distance of Taurus. In the gas-depleted disc, we see similar amounts of clumping in the outer disc, but with dust densities up to a factor of ten higher, and some signs that the streaming instability might be developing at 3 au as well. Thus as discs age we predict that the streaming instability lends itself more easily to observational confirmation.

### 3.3. Possibilities for grain growth

Grain growth for large numbers of colliding particles is determined by the Smoluchowski equation (Smoluchowski 1916), and the rate of growth is proportional to the product of the densities of the colliding particles. The streaming instability discussed above causes the dust density to increase significantly over the enhancement due solely to sedimentation. Furthermore, the streaming instability is also known to reduce the relative velocity between particles (Balsara et al. 2009) which is also known to increase the probability that two particles will stick during a collision, rather than destroy each other (Blum & Wurm 2008). Consequently, it is worthwhile to examine the implications this density enhancement has for the growth of dust grains. This is the first time such a calculation has been attempted based on simulated data. The sticking probabilities used here are the best we could cull from the literature.

The growth of dust in protostellar discs has been studied extensively through the use of semi-analytic models (Suttner & Yorke 2000; Dullemond & Dominik 2005, etc). In these models, grain-grain collisions arise through parameterizations of grain turbulence and sedimentation. In the absence of fragmentation, grain growth is extremely rapid, with centimeter and meter-sized grains forming on time-scales of a few thousand years and micron-sized grains depleting on a similar time-scale. Such a rapid depletion of the small grains is inconsistent with the infrared emission that is seen in discs a million years old. Dullemond & Dominik (2005) posited that significant levels of grain fragmentation were required in order to maintain a population of small grains to explain the infrared emission. They found that even with high levels of fragmentation, they could get large grains to form.

A common feature in these simulations is that the collision velocities between dust grains arise from semi-analytic models of disc turbulence, such as those studied by Völk et al. 1980. In these models, the average collision velocity between grains arises from perturbations to laminar motions arising from turbulent eddies of different sizes and the differences in the strength of the coupling of the grains to gas motions within these eddies. The dust-gas interaction in these models is usually considered only in the limits where the friction time-scale is much less or much larger than the eddy turnover time, which is comparable to the orbital time-scale. As we discussed in Section 3.2, the streaming instability is most sensitive to the case where the friction time-scale is comparable to the orbital time-scale, so further clarity of the dust-dust collision velocities in this regime is desirable. Furthermore, it is not clear whether a semi-analytic model will capture the intermittency of the gas turbulence, or treat the interaction of eddies. Our simulations do include these physics, and thus we can estimate a growth rate for dust and compare it to the results that arise from semi-analytic models.

The collision rate per unit volume between dust of family i with dust of family j is given by $R_{ij}(\mathbf{x}) = n_i(\mathbf{x}) n_j(\mathbf{x}) \sigma_{ij}(\mathbf{x}) \mathrm{v}_{ij}(\mathbf{x}) S(\mathrm{v}_{ij}(\mathbf{x}))$, where $n_i(\mathbf{x})$ and $n_j(\mathbf{x})$ are the particle densities at grid zone located at $\mathbf{x}$, $\sigma_{ij}(\mathbf{x})$ is the interaction cross-section for particle i with particle j, $\mathrm{v}_{ij}(\mathrm{x})$ is the collision speed and $S(\mathrm{v}_{ij}(\mathrm{x}))$ is a sticking coefficient (see, for example, Eq. 32 of Suttner & Yorke 2001). The time-scale for a dust grain of family i to collide with a number of dust grains of family j with total mass equal to the mass of grain i is thus $\tau_{ij}(\mathbf{x}) = n_i(\mathbf{x}) m_i \big/ \left[ m_j R_{ij}(\mathbf{x}) \right]$. In principle, if every collision resulted in the particles sticking together (i.e. if $S$=1 for all collision velocities), then this would also be the time-scale for grains of family i to grow by accreting grains from family j. However, only those collisions that take place with a small-enough relative velocity result in the sticking of the two grains.

In practice, we calculate this time-scale from our distribution of superparticles. To do this, we use the same triangular-shaped cloud interpolation strategy that we use to treat the dust-gas interaction, as detailed in Paper I. If $W_{TSC}(\mathbf{x}-\mathbf{x}_k)$ is the interpolation kernel for a particle at location $\mathbf{x}_k$ to the grid zone at $\mathbf{x}$, $\mathrm{dV}(\mathbf{x})$ is the volume of the zone at $\mathbf{x}$, and $N_k$ is the number of dust grains that are contained in the superparticle k, then the density of dust family i at any point in the simulation is given by

$$n_i(\mathbf{x}) = \sum_{k \in i} W_{TSC}(\mathbf{x} - \mathbf{x}_k) N_k \mathrm{dV}(\mathbf{x})$$

and the collision time is given by

$$\tau_{ij}(\mathbf{x}) = \frac{m_i}{m_j} \frac{\sum_{k \in i} W_{TSC}(\mathbf{x}-\mathbf{x}_k) N_k \mathrm{dV}(\mathbf{x})}{\sum_{k \in i}\sum_{p \in j} W_{TSC}(\mathbf{x}-\mathbf{x}_k) W_{TSC}(\mathbf{x}-\mathbf{x}_p) \sigma_{ij} \left|\mathbf{v}_k - \mathbf{v}_p\right| S(\mathbf{v}_k - \mathbf{v}_p) N_k N_p \mathrm{dV}(\mathbf{x})^2} \quad (6)$$

Note that this formulation also allows us to consider the case where i=j, i.e. we can consider collisions with grains of the same family. If $\tau_{ij}$ is much smaller than the orbital period for any particular families i and j, then we would expect significant amounts of momentum to be exchanged between those families and our collisionless approximation for the dust breaks down. If the grains also stick together during these collisions, we would further expect extremely rapid grain growth. If $\tau_{ij}$ is much larger than the orbital time, then grain collisions are very infrequent and grain growth is slow, even if the grains always stick together during a collision. The intermediate regime, where $\tau_{ij}$ is comparable to the orbital time, is of great interest as the rate of momentum exchange between grains is small enough to justify our collisionless approximation, but the potential for rapid grain growth exists.

We use Equation (6) to calculate the time-scale for dust particles to interact with their own mass with other dust families of different sizes, assuming perfect sticking ($S$=1). We plot the minimum and mean of these interaction time-scales in Figure 10. In each of the panels in Figure 10, the black stars denote the time-scale for 2.5 cm grains to interact with an equal mass of grains of different sizes from the x-axis. Similarly, the orange diamonds are the interaction time-scale for 2.5 mm grains with grains of 2.5 mm or smaller, the blue crosses are the interaction time-scales for 250 μm grains with grains of equal size or smaller, the green triangles are the interaction time-scales for 25 μm grains with an equal mass of grains of equal size or smaller, and the red square is the time-scale for 2.5 μm grains to interact with other 2.5 μm grains. The left column of Figure 10 shows the minimum interaction time-scales in the simulations at 1 au, 3 au, 10 au, 30 au and 100 au, from top to bottom respectively. The right column similarly shows the mean interaction time-scale.

A general trend that can be seen in Figure 10 is that large dust grains tend to collide with grains of the same size on a time-scale as short as a few orbits beyond 3 au, and on time-scales of tens of orbits at radii of 3 au or less. Smaller grains collide with themselves on time-scales less than an orbit. The enhanced midplane density from sedimentation has reduced the collision time-scale for the large dust significantly. The streaming instability in the midplane has further enhanced the density of sedimented particles and led to another reduction of about one order of magnitude in the collision time-scale.

In Figures 10 we see that the sedimentation of large grains leads to an increased predominance of large grains colliding with other large grains, rather than large grains colliding with smaller grains. For instance, the minimum time-scale for 2.5 mm dust at 100 au to interact with its own mass of 250 μm dust is 18.4 orbits. The same 2.5 mm dust will interact with itself on time-scales as low as 2.2 orbits. Similarly, the minimum time-scale for 2.5 cm grains at 10 au to interact with 2.5 mm grains is 307 orbits, while the minimum time-scale for 2.5 cm grains to interact with themselves is 10.5 orbits. Conversely, interior to 3 au grains primarily collide with smaller grains, as seen in Figures 10e and 11e. As a result, we would expect from this argument that if we had perfect sticking in every collision, then grains would grow predominantly through collisions with similar-sized grains in the outer disc – the scenario that forms the BCCA structure for fractal grains – while growing primarily through accreting smaller grains in the inner disc, a scenario predicted to form much more compact grain structures.

We also see from Figures 10 that the smaller grains – which do not experience significant amounts of sedimentation – will primarily collide with smaller families of dust, and would thus likely grow as compact grains.

In Figure 11, we show the interaction time-scales at 30 au in the midplane for four typical grain-grain collisions. In Figure 11a, we show the collisions of 2.5 mm grains with an equal mass of 2.5 μm grains. We see that this typically occurs on time-scales of a few tens of orbits, and that there are no strong concentrations where the collision rate is significantly enhanced. In Figure 11b, we show the collisions of 2.5 cm grains with an equal mass of 2.5 mm grains, and in Figure 11c we show the collisions of 2.5 mm grains with 2.5 mm grains. In both of these panels, we see that the lowest collision times are found only in a few highly concentrated regions. In Figure 11d, we show the collision time-scales for 2.5 cm dust with itself. Here again, we see that the collision rate is maximal in only a very small volume in the midplane. It is only in these regions where the density is enhanced by the streaming instability that significant enhancements in the growth rate can be found.

The interaction time-scales above are the time-scales for the collisions of particles with other grains of their own size, or for collisions of the grains of a certain size with an equal mass of grains of a smaller size. There are many possible outcomes for each collision, though. Particles can stick together, particularly if the collision velocity is low. The grains can also scatter off each other, or they can fragment and break apart. Blum & Wurm (2008) exhaustively catalogue the current status of dust collision experiments that test the collision behaviour of dust in laboratory settings. However, the collision velocities from their disc model that they use to guide their experimental setups tend to be smaller than the typical collision velocities that we find in our simulations, particularly for the smallest dust. In their disc models, sub-centimetre-sized grains have collision velocities less than 1 cm $s^{-1}$; the vigorous MRI-driven turbulence in our simulations, however, leads to collision velocities between dust grains that are typically a few m $s^{-1}$. Thus, there is some uncertainty as to what the appropriate velocity threshold for fragmentation should be for our simulations. However, as a general rule these experiments show that fragmentation predominates over coagulation when the collision velocity exceeds 1-2 m $s^{-1}$. To assess the impact this has in the context of our simulations, we repeat the calculation of Equation (6), but now use a sticking coefficient that is more consistent with experiments. Blum & Wurm (2000) found that the rate of grain destruction exceeded the rate of coagulation when the collision velocities were greater than ~1.9 m $s^{-1}$. For the purposes of this calculation, we will simplify this slightly and use $S(\mathbf{v}_k\text{-}\mathbf{v}_p) = 1$ for $|\mathbf{v}_k\text{-}\mathbf{v}_p| < 1.9$ m $s^{-1}$, and $S(\mathbf{v}_k\text{-}\mathbf{v}_p) = 0$ otherwise. In other words, we only add a contribution to the collision rate if the collision velocity is less than the critical velocity of 1.9 m $s^{-1}$.

This restrictive condition has a large impact on the growth rates that we estimate from our simulations. We plot in Figures 12 the growth rate for grains of different sizes and different radii, analogous to Figures 10. Similarly, in Figure 13 we plot the growth rates in the midplane at 30 au for 2.5 mm grains growing through collisions with 2.5 μm grains, for 2.5 cm grains to grow via collisions with 2.5 mm grains, for 2.5 mm grains to grow through collisions with 2.5 mm grains, and for 2.5 cm grains to grow through collisions with other 2.5 cm grains. The minimum growth time-scales have been reduced by a factor of ~100-1000 to the collision time-scales we found previously. At 10 au, for instance, the minimum growth time-scale for 2.5 cm grains through collisions with other 2.5 cm grains is 3100 orbits; the minimum growth time-scale for 2.5 cm grains with 2.5 mm grains is 2.71 x $10^5$ orbits. The minimum growth time-scale for 2.5 cm grains with 2.5 μm grains is a little faster, at 2.93 x $10^3$ orbits. The other panels in Figure 12 show that these trends persist for other dust sizes and at other radii in the disc; the growth rate for grains is substantially depressed when we consider just the grain collisions with collision velocities less than 1.9 m $s^{-1}$.

The large collision velocities that we find will not necessarily prevent any grain growth from occurring. First of all, one of the models considered by Dullemond & Dominik (2005) considered fragmentation in addition to grain growth. In this model, those authors found that while the majority of grains were fragmented in collisions before they could grow to centimeter sizes, there was a fraction of the grains that escaped undergoing a fragmentation event that grew to meter sizes over the course of a few hundred thousand years. Secondly, highly porous grains may undergo compactification in collisions, rather than fragmentation or cratering. Furthermore, the smaller masses of fractal grains for a given size reduces the collision energy, allowing collisions to occur at higher velocities before the grains begin to fragment. Finally, the collision time-scales that we find are at best comparable to the lifetime of the clumps they are in, and in many cases are longer. As a result, dust grains may be swept up in a clump, undergo some degree of coagulation over the course of a few orbits, and then disperse back into the ambient disc where fragmenting collisions are more rare.

The time-scale for momentum transfer between dust particles in the same family that we extract from Figures 10 and 11 generally correspond to at minimum tens to hundreds of timesteps per collision. Thus, our approximation that the gas is collisionless is verified. The only exception comes for the smallest dust at 0.3 and 1 au. The collision time-scale for the 2.5 μm dust at 0.3 au is approximately one-third of a timestep, and the time-scale for 25 μm dust at 0.3 au is approximately one collision every three timesteps. The 2.5 μm dust at 1 au has approximately one collision every two timesteps. We would thus expect dust self-collisions to have some small role for these dust families in the inner disc. However, the momentum exchange rate between 2.5 μm dust and the gas is on the order of $10^4$ larger, and thus should dominate any dust pressure.

**4) Summary**

The streaming instability is pervasive throughout sedimented dusty discs, and can lead to significant enhancements in the densities of large grains. We have catalogued the sedimentation and development of streaming instabilities throughout a prototypical protostellar disc. We find that:

1) Dust of different grain sizes sediment to different degrees, with larger dust sedimenting to a greater extent than smaller grains and more sedimentation occurring in the outer disc. This effect can become quite pronounced, producing dusty sub-layers with thicknesses of 1-2 au. Such layers should be distinguishable from thicker discs via observations by ALMA at submillimeter wavelengths for discs at the distance of Taurus.

2) We show that the streaming instability produces clumps in 250 µm grains, 2.5 mm grains and 2.5 cm grains that are on the order of a few au across in the outer disc, especially in gas-depleted systems like transitional discs. ALMA at 350 µm should be able to resolve features as small as 0.7 au at a distance of 140 pc, and can thus potentially detect these dust clumps when viewing face-on discs.
3) The scale height of the sedimented dust approaches a power-law relationship with dust size for compact grains, with a slope slightly steeper than -1/2. Highly porous fractal grains have a scale height that is largely independent of grain radius.
4) The combination of sedimentation and the streaming instability leads to dust-to-gas ratios of ~1 over the course of a few orbits, and increase of ~100 over our initial conditions.
5) The enhanced grain densities that are produced by sedimentation and the streaming instability in turn lead to a reduced collision time-scale for both large grains with other large grains, as well as for large grains with small grains, leading to the possibility of enhanced dust growth rates.
6) The collision velocities of the dust are on the order of a several metres per second to a few tens of metres per second, significantly higher than that assumed by many semi-analytic models of dust growth in protostellar discs. As a result, while the collision rates between dust particles can be significantly enhanced, the most likely result of the majority of such collisions is likely to be grain destruction, rather than grain growth.

Table 1: List of simulation parameters.

| Radius (au) | T (K) | $c_s$ (km $s^{-1}$) | $H_{gas}$ (au) | Σ (g $cm^{-2}$) | $\rho$ (g $cm^{-3}$) | $t_{orbit}$ (yr) |
|---|---|---|---|---|---|---|
| 0.3 | 365.0 | 1.33 | 0.0146 | $1.04 \times 10^{4}$ | $2.67 \times 10^{-8}$ | 0.232 |
| 1.0 | 232.6 | 1.06 | 0.0711 | $1.70 \times 10^{3}$ | $9.02 \times 10^{-10}$ | 1.41 |
| 3.0 | 148.1 | 0.845 | 0.295 | $3.28 \times 10^{2}$ | $4.17 \times 10^{-11}$ | 7.35 |
| 10.0 | 95.85 | 0.680 | 1.44 | 53.8 | $1.41 \times 10^{-12}$ | 44.7 |
| 30.0 | 69.12 | 0.577 | 6.37 | 10.4 | $6.13 \times 10^{-14}$ | $2.32 \times 10^{2}$ |
| 100.0 | 50.24 | 0.492 | 33.0 | 1.70 | $1.94 \times 10^{-15}$ | $1.41 \times 10^{3}$ |
| 150.0 | 42.72 | 0.454 | 56.0 | 0.925 | $6.24 \times 10^{-16}$ | $2.60 \times 10^{3}$ |
| 300.0 | 32.37 | 0.395 | 138 | 0.327 | $8.95 \times 10^{-17}$ | $7.35 \times 10^{3}$ |

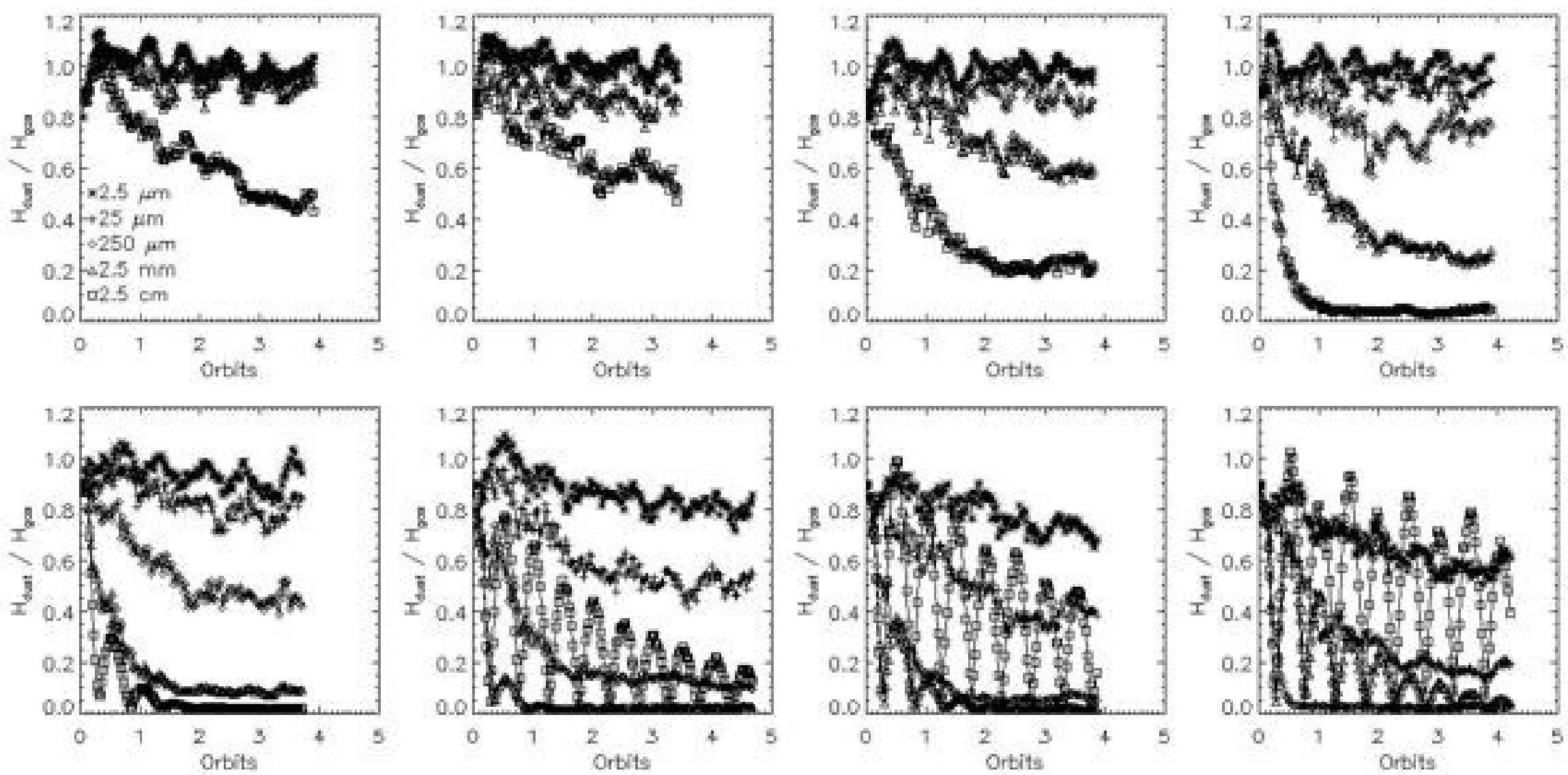

Figure 1. Scale height of spherical dust grains relative to the scale height of gas for dust of various sizes as a function of time in our simulations, at several orbital locations.

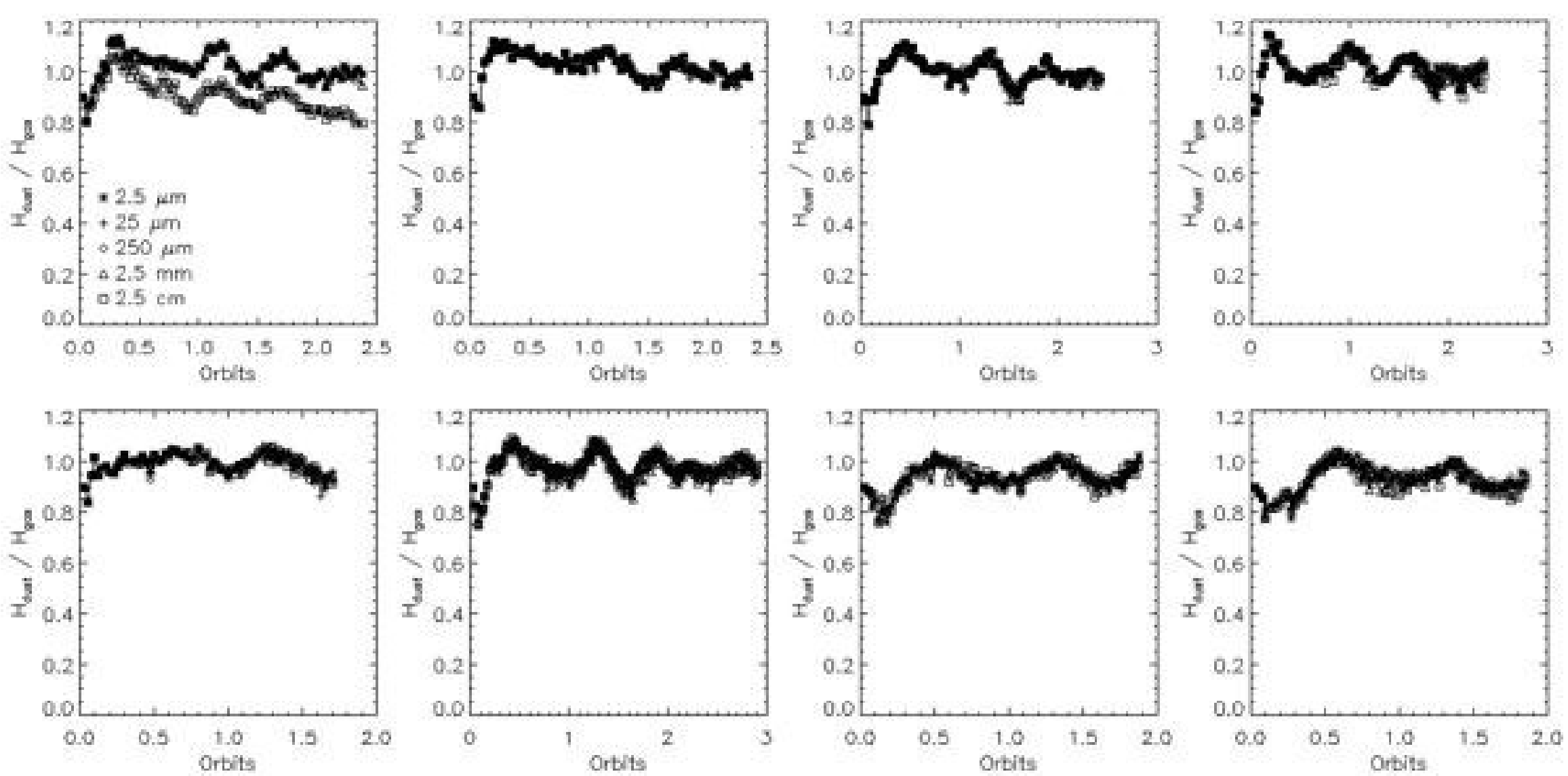

Figure 2. Dust scales as a function of time for fractal BCCA grains.

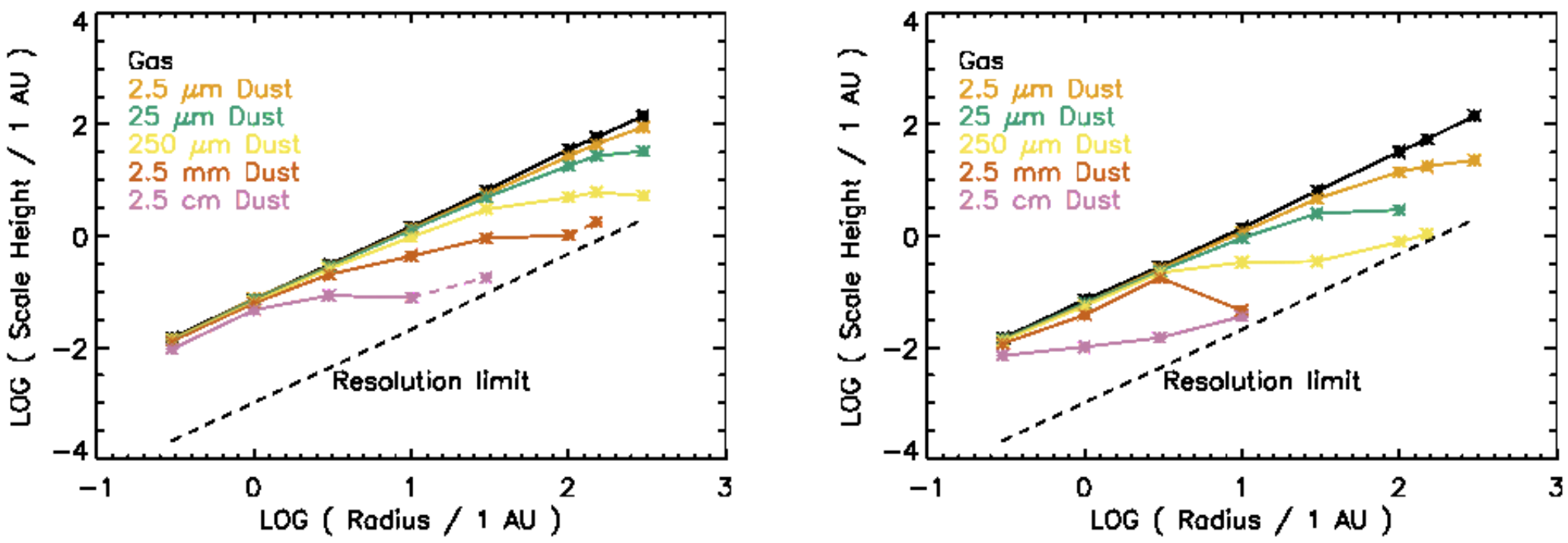

Figure 3. Dust and gas scale heights as a function of radius in the disc. (Top) Minimum-mass solar nebula. (Bottom) Disc with the gas density depleted by a factor of 10.

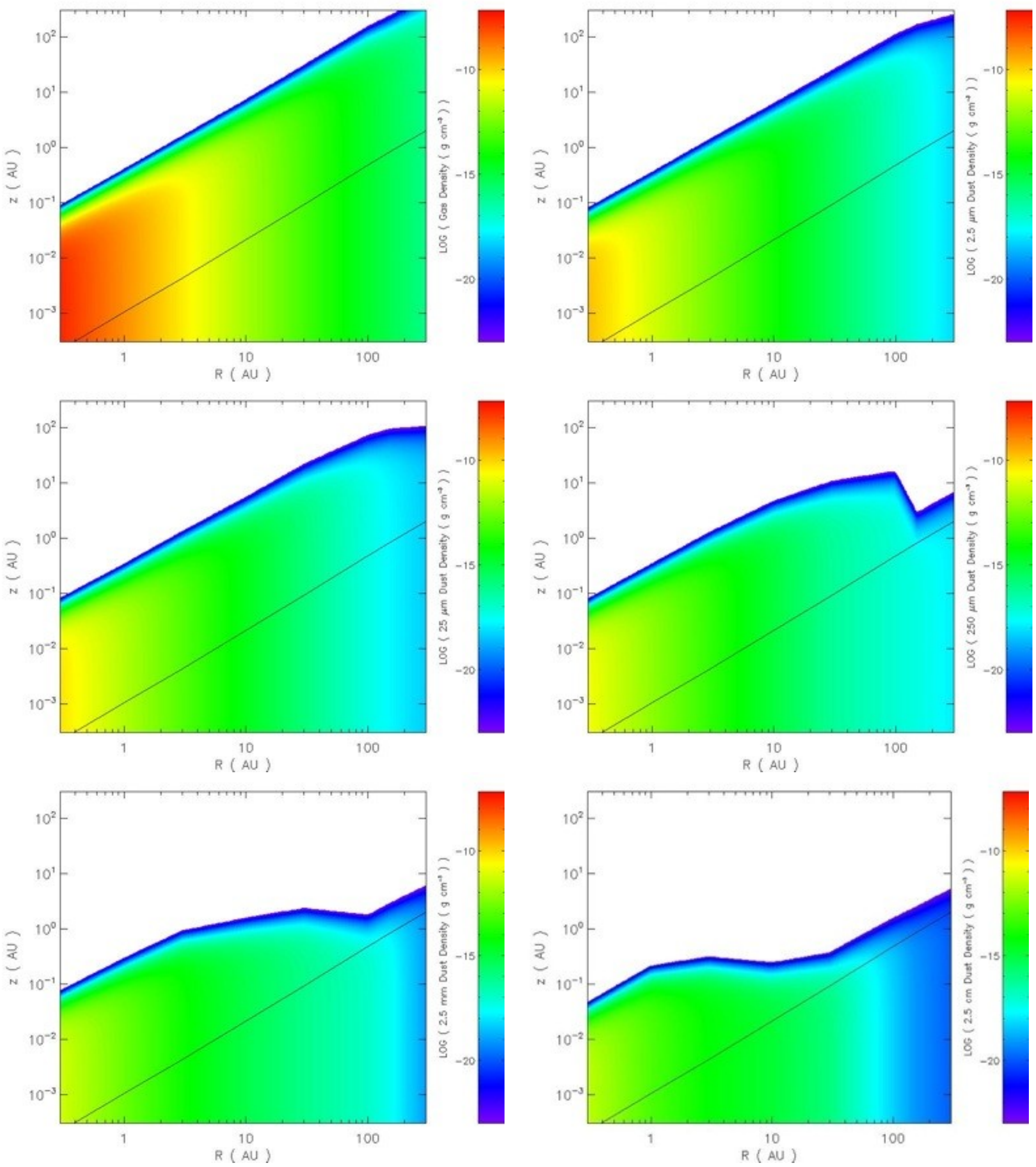

Figure 4. Gas and dust density interpolated from the final midplane density and scale heights from our simulations, over the entire R-z structure of the disc with logarithmic scaling of the axes. The solid line is marks the resolution of our simulation, and effectively places a lower limit on the scale heights we can achieve. The top left panel shows the gas density; the top right panel shows the density of the 2.5 μm grains; the middle left panel shows the 25 μm grain density; the middle right panel shows the 250 μm density; the lower left panel shows the 2.5 mm grain density, and the lower right panel shows the 2.5 cm grain density.

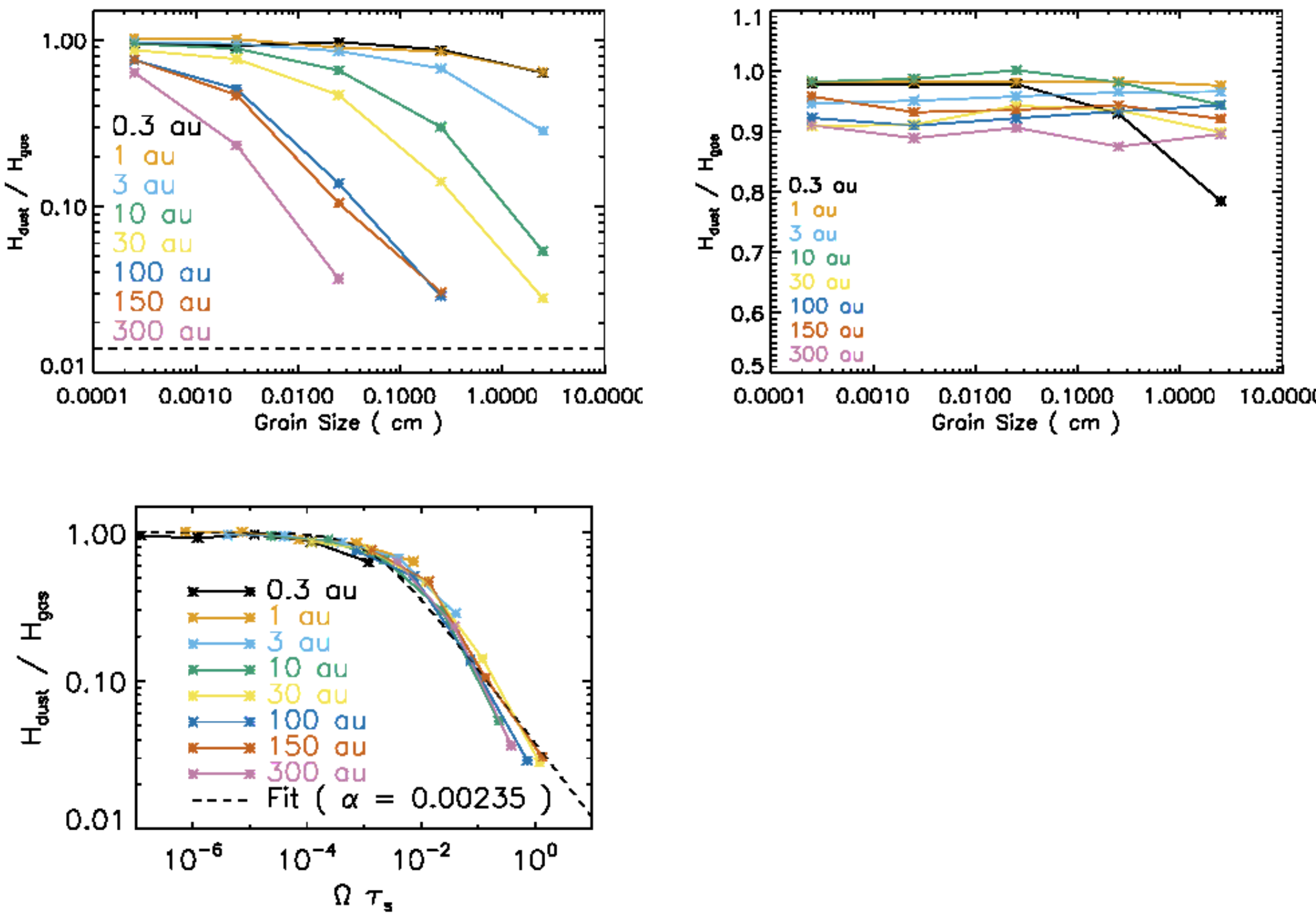

Figure 5. Ratio of the scale height of dust to the scale height of gas in the disc when the dust layer reaches a steady state. (a) Scale heights for spherical grains as a function of grain size. The largest grains at 100 au and beyond oscillate through the midplane and do not reach a steady state, and so are not shown here. The horizontal dashed line is the resolution limit for our simulations. (b) Scale heights for BCCA fractal grains. Note that the vertical axis is linear, not logarithmic. (c) Scale heights for spherical grains, as a function of dimensionless stopping times.

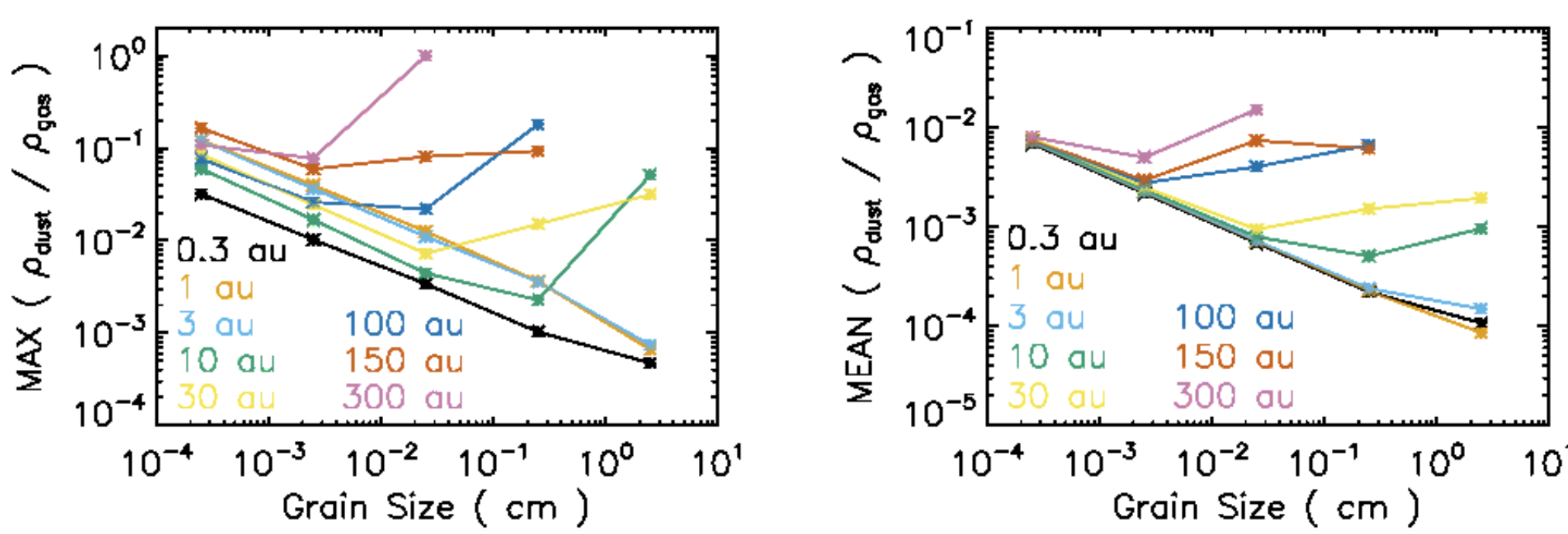

Figure 6. (a) Maximum dust-to-gas ratio in the disc for spherical grains of different sizes at each radial station. The dust that oscillates rather than settles is not shown. (b) Mean dust-to-gas ratio in the midplane of the disc for grains of different sizes at each radial station.

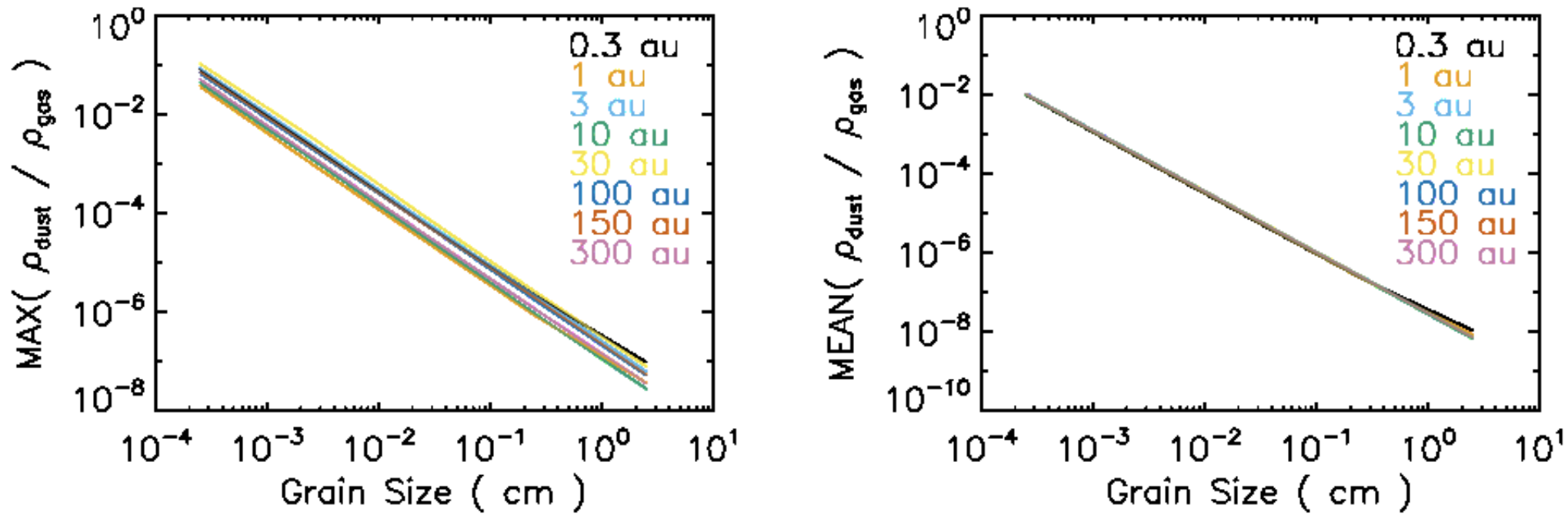

Figure 7. (a) Maximum dust-to-gas ratio in the disc for fractal grains of different sizes at each radial station. (b) Mean dust-to-gas ratio in the midplane of the disc for grains of different sizes at each radial station.

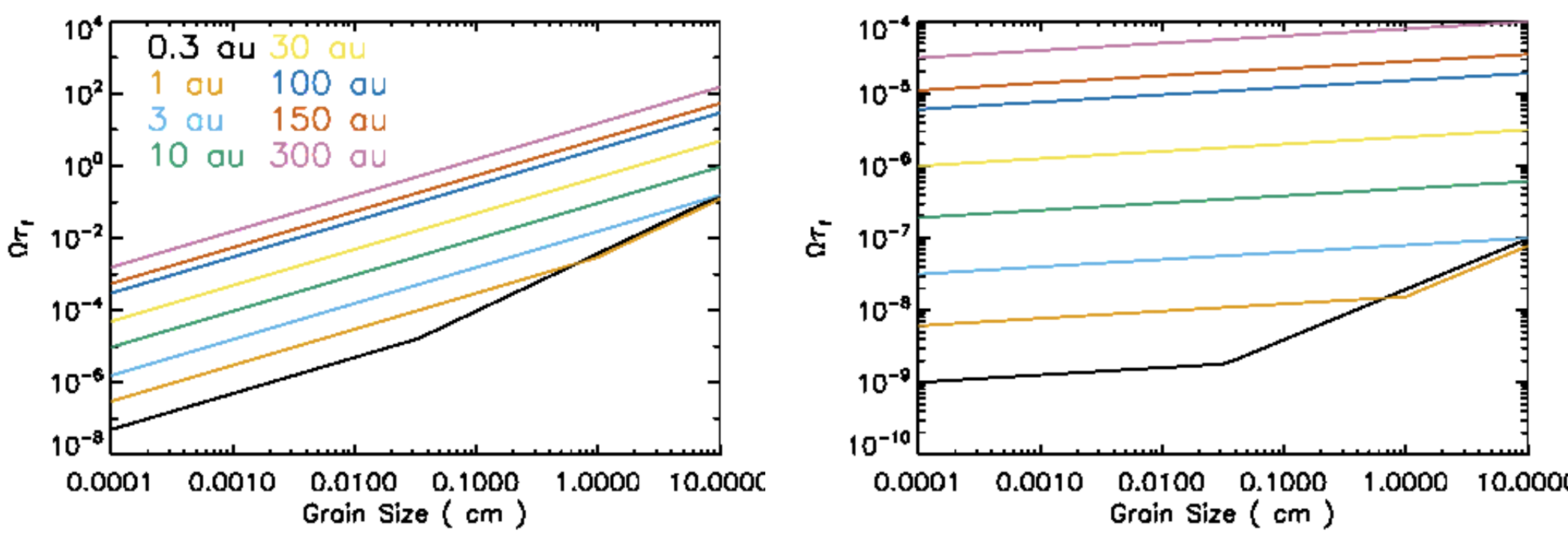

Figure 8. (a) Dimensionless stopping time $\Omega\tau_f$ as a function of grain size for the different radial stations for spherical grains. (b) Dimensionless stopping time $\Omega\tau_f$ as a function of grain size for the different radial stations for BCCA grains.

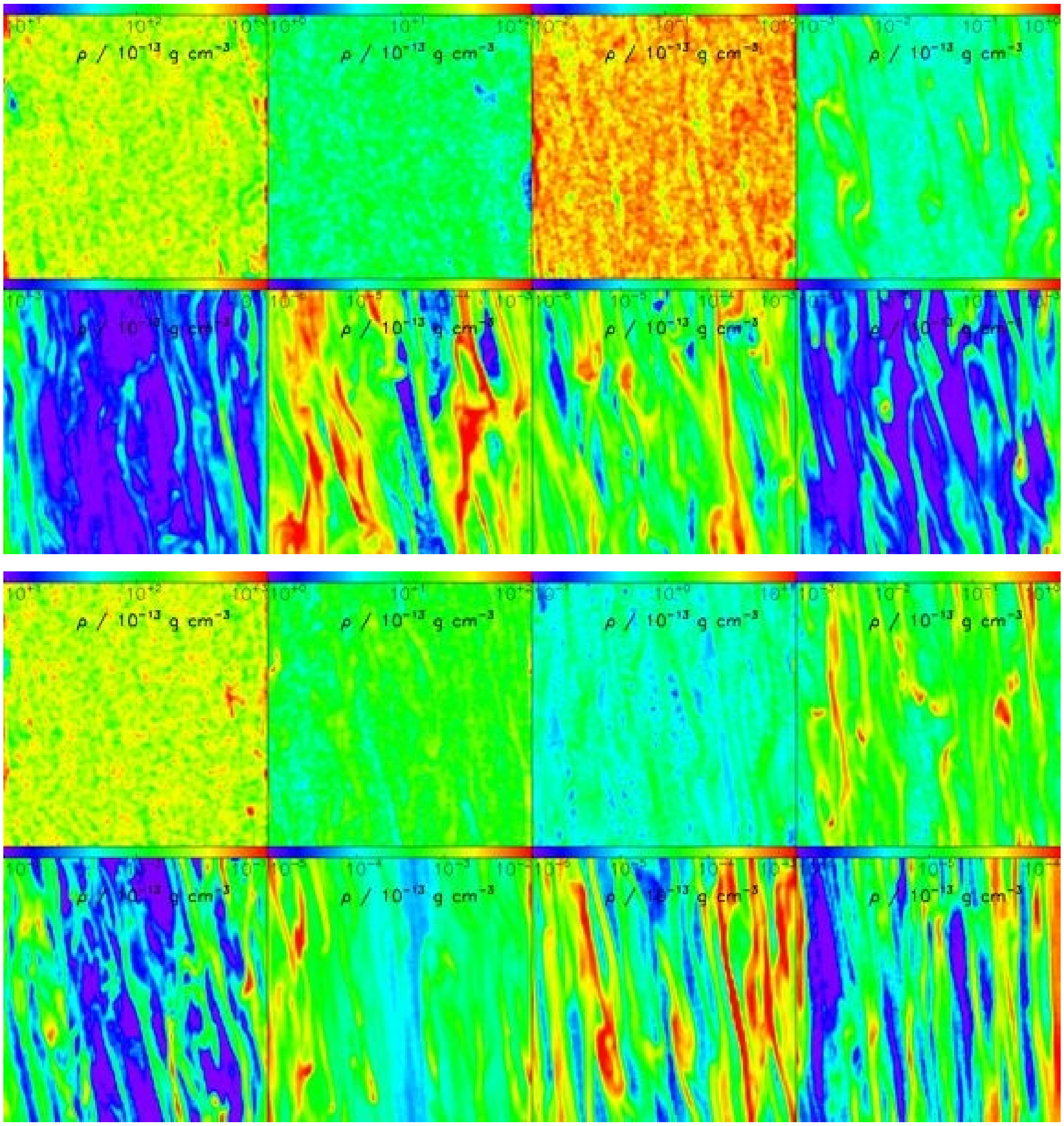

Figure 9. (First row) The total midplane density for 250 μm – 2.5 cm grains at 0.3 au, 1 au, 3 au and 10 au in a MMSN. (Second row) The total midplane density for 250 μm – 2.5 cm grains at 30 au, 100 au, 150 au and 300 au in a MMSN. (Third row) The total midplane density for 250 μm – 2.5 cm grains at 0.3 au, 1 au, 3 au and 10 au in a gas-depleted disc. (Fourth row) The total midplane density for 250 μm – 2.5 cm grains at 30 au, 100 au, 150 au and 300 au in a gas-depleted disc.

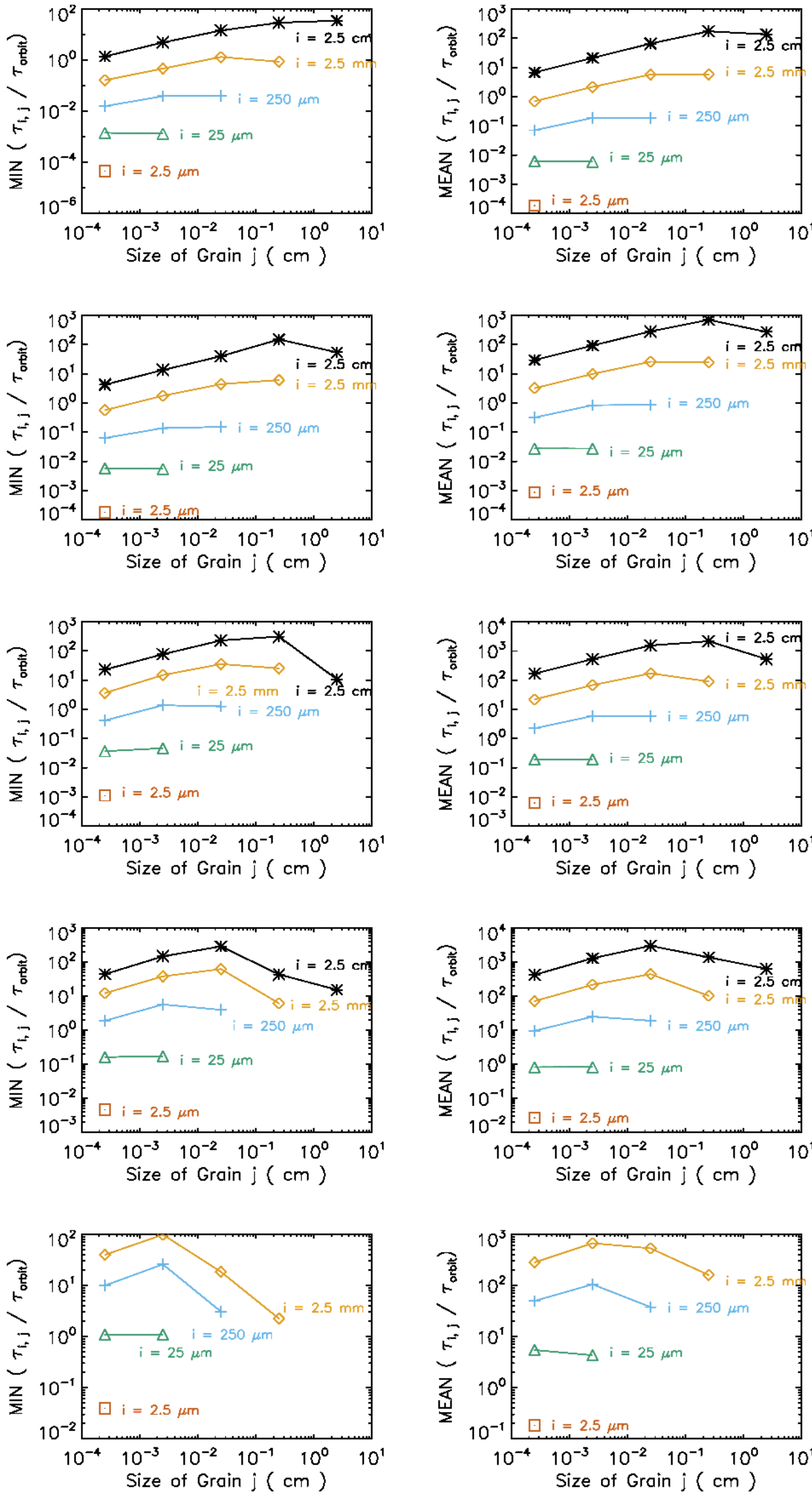

Figure 10. (Left column) Minimum collision time-scales in the midplane for grains of family i to collide with an equal mass of grains from family j, where i is labelled next to the line in question and j is the x axis. (Right column) Mean time for grains of family i to collide with an equal mass of grains from family j. For both columns, from top to bottom are the data for spherical grains at 1 au, 3 au, 10 au, 30 au and 100 au, respectively. At 100 au, the 2.5 cm grains oscillate about the midplane instead of settling, and so are not shown.

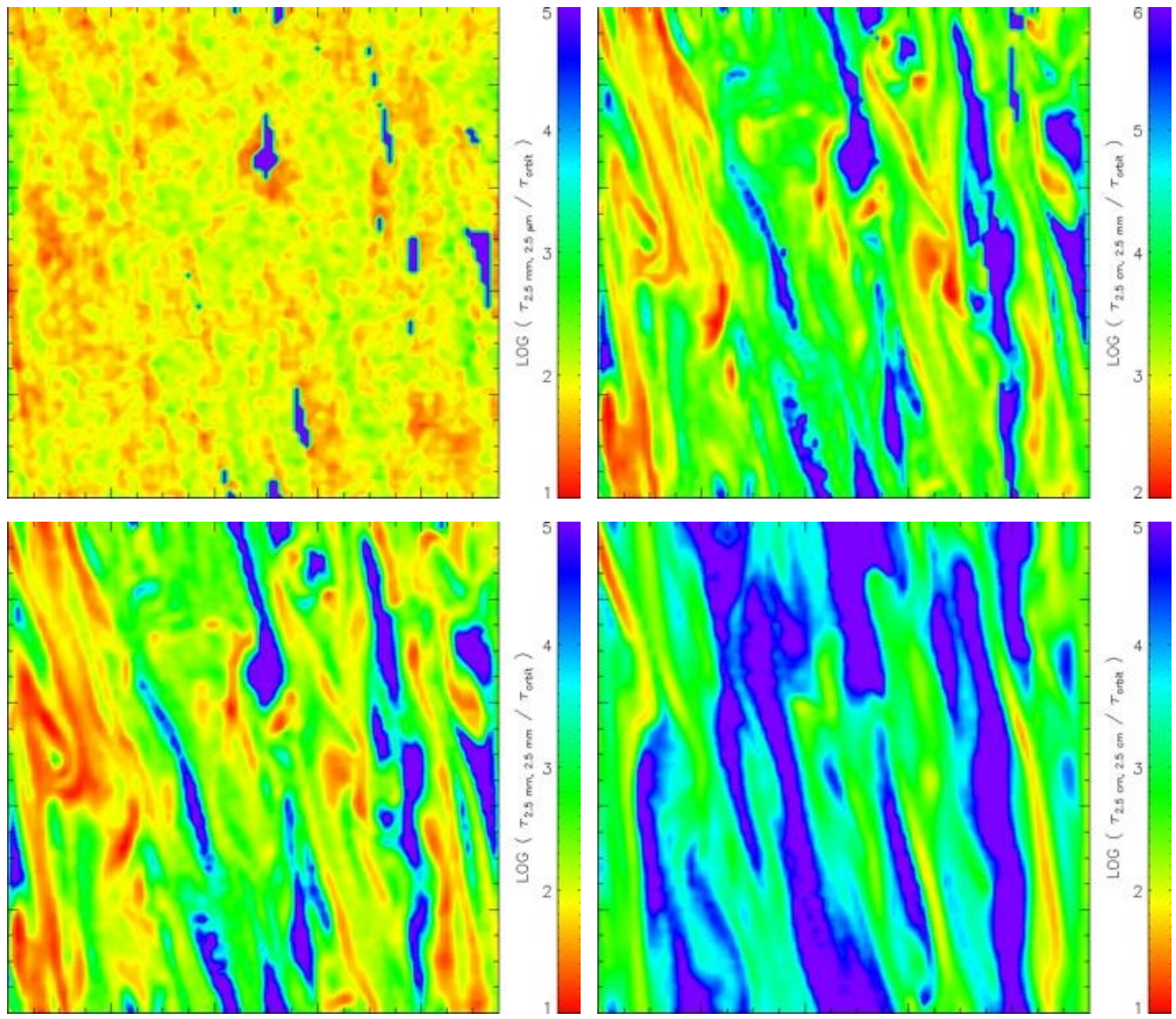

Figure 11. (a) Time-scale for 2.5 mm grains at 30 au to collide with an equal mass of 2.5 μm grains. (b) Time-scale for 2.5 cm grains at 30 au to collide with an equal mass of 2.5 mm grains. (c) Time-scale for 2.5 mm grains at 30 au to collide with 2.5 mm grains. (d) Time-scale for 2.5 cm grains at 30 au to collide with 2.5 cm grains.

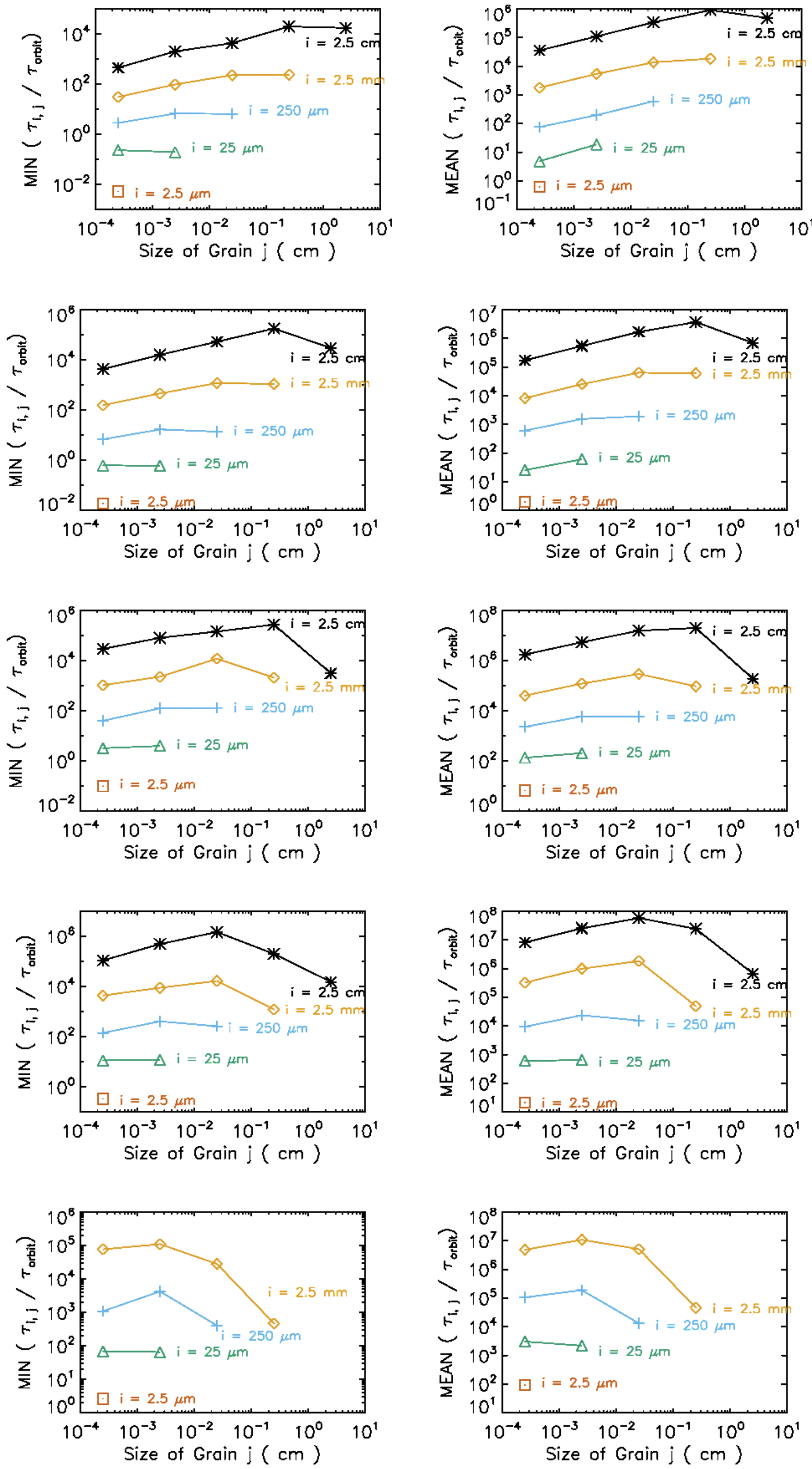

Figure 12. The time-scales for grains of various sizes to grow through collisions with grains of other sizes. Grains will only grow if their mutual collision velocities are less than 1.9 m $s^{-1}$ (see main text). (Left column) Minimum time-scales in the midplane for grains of family i to grow via collisions with an equal mass of grains from family j, where i is labelled next to the line in question and j is the x axis. (Right column) Mean time for grains of family i to grow via collisions with an equal mass of grains from family j. For both columns, from top to bottom are the data for spherical grains at 1 au, 3 au, 10 au, 30 au and 100 au, respectively. At 100 au, the 2.5 cm grains oscillate about the midplane instead of settling, and so are not shown.

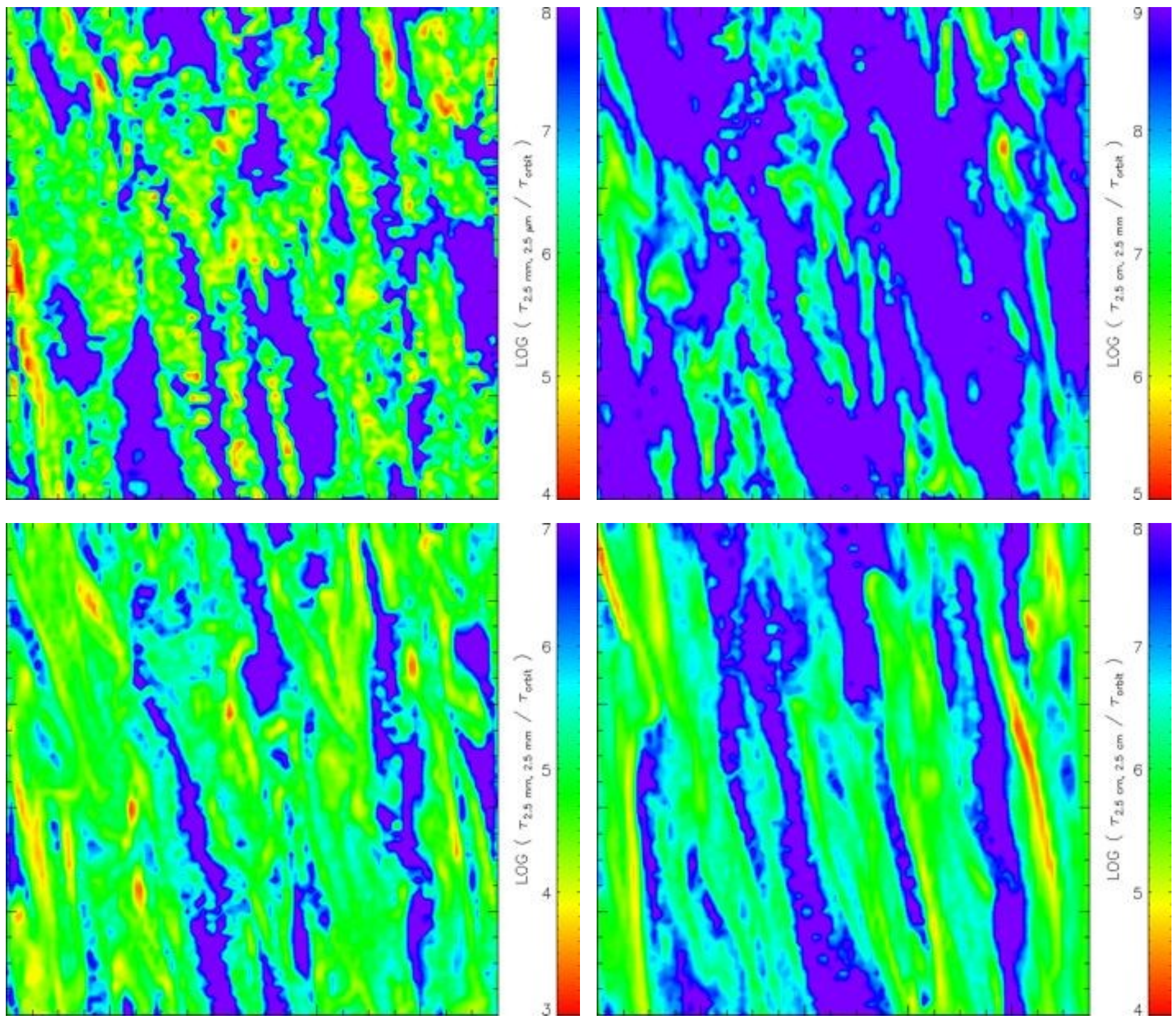

Figure 13. (a) Time-scale for 2.5 mm grains at 30 au to grow through collisions with 2.5 μm grains. (b) Time-scale for 2.5 cm grains at 30 au to grow via 2.5 mm grains. (c) Time-scale for 2.5 mm grains at 30 au to grow via 2.5 mm grains. (d) Time-scale for 2.5 cm grains at 30 au to grow via 2.5 cm grains.